\documentclass[a4paper]{article}
\usepackage{graphics}

\def\vybiral{Vyb\'{\i}ral}%
\def\voracek{Vor\'{a}\v{c}ek}%
\def\horak{Hor\'{a}k}%
\def\hradec{Hradec Kr\'{a}lov\'{e}}

\title{"Autothixotropy" of Water - \\
	an Unknown Physical Phenomenon}
\author{Bohumil \vybiral\footnote{Email: Bohumil.Vybiral@uhk.cz}\ \ and Pavel \voracek\footnote{Address: Lund Observatory, Box 43, SE-221 00 Lund, Sweden}\\
	\\
	Department of Physics, Pedagogical University,\\
	n\'{a}m. Svobody 301, CZ-501 91 \hradec, Czechia}
\date{June 10, 2003}
\begin{document}

\maketitle

\begin{abstract}
A complex of until now unknown phenomena ongoing in water was discovered in laboratory experiments, where it
made impossible gravimetric measurements with the necessarily extreme precision. This behaviour of water,
which we call \emph{autothixotropy}, was the issue of the presented experimental research. We are also proposing
a possible explanation.
\end{abstract}

\begin{verse}
\textbf{Motto:} \emph{The average lifetime of a hydrogen bond between water molecules at room temperature is three picoseconds.}
(The actual - generally accepted - scientific opinion as presented by David Labrador in \emph{Scientific American}, September 2002, p. 38.)
\end{verse}

\section{Laboratory Observations}

During 1978-1986 a series of measurements were performed (\vybiral, 1987 and 1989) in order to verify the gravitational
law in fluids as deduced by \horak\ (1984).  In 1978 a disturbing phenomenon was observed in the measurements which
compelled the author to use another method.

\begin{figure}[bt]
 \begin{center}
  \resizebox{!}{12cm}{\includegraphics{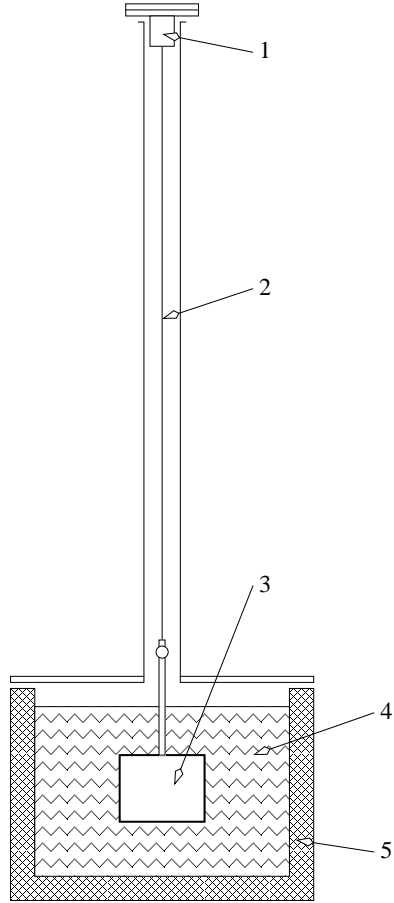}}
  \caption{Schematic drawing of the experimental device:
	1 - fine-turning hanger; 2 - phosphor-bronze filament (cross-section 0.20 mm x 0.025 mm, length 0.811~m);
	3 - stainless steel plate  (b.60 mm, h.50 mm, t.1 mm); 4 - distilled air-free water;
	5 - heat-insulated closed sterilized vessel (inner diameter 150 mm)}
  \label{stroj}
 \end{center}
\end{figure}

In a private communication between the authors, the new - then disturbing - phenomenon was discussed and a series of
experiments focusing on this phenomenon were proposed.  In the Department of Physics of Pedagogical University at
\hradec, the original experimental device was reconstructed with the purpose of creating the best conditions
for the observation of the phenomenon (see Fig.~\ref{stroj}).

After objects immersed in the water have been at rest for one or several days,
seven qualitatively different phenomena can be observed:

\begin{description}
\item[(1)] When the hanger is rotated by a certain angle, the plate immersed in the water remains in almost the same
position, in spite of the twisting tension arising in the thin filament. (Nevertheless, a "creep" toward a new
neutral position can be observed after some time (days or weeks).)  When a certain critical angle is reached, the
plate will rotate - relatively quickly - into a new neutral position given by the angle of rotation of the hanger.  
(When a smooth-surfaced cylinder, which can rotate around its own axis, was used instead of the plate, this phenomenon was not
observed.)

\item[(2)] Another phenomenon, though - in the clarity of appearance - weaker than the phenomenon mentioned above, can
simultaneously be observed:  a quick rotation of the plate in the direction of the rotation of the hanger; nevertheless,
the angle of the quick rotation is one or two orders of magnitude less than the angle in phenomenon (1).

(Using a one generation older device, the strongly dampened oscillations could be observed around the new equilibrium
position (i.e. rest-position with twisting tension in the filament) after the mechanical pulse had been given to the
plate. Such oscillations were directly observed by \voracek\ in a very rough experiment performed in March 1991 (not
published) and (after a personal report) discovered by \vybiral\ when reanalyzing observational material from 1978).

\item[(3)] The critical angle of rotation in phenomenon (1) is dependent on the period of time the water has been at
rest.  This angle increases with time, starting from virtually zero.  (The critical angle can reach values of
several tens of degrees!)

\item[(4)] If the plate is only partially immersed, the critical angle is significantly greater than when it is immersed
completely.

\item[(5)] In case of partial immersion, the phenomenon analogous to (2) is much more prominent.

(Both phenomena (4) and (5) are time-dependent as described in (3) despite the time-invariance of the surface tension
which we also tested.)

\item[(6)] If the water is stirred after having been at rest for some time (several days), the critical angle increases from
zero quicker (when being again at rest) than when new "fresh" water is used.

\item[(7)] The critical angle is significantly augmented and the phenomenon appears earlier if the (distilled) water is
boiled (thus deaerated) before the experiment is started.
\end{description}

Regarding the convincing prominence of observed phenomena, we consider them being significant enough in order to claim
they are real, despite the performed experiments were only qualitative. (Our attempts in carrying out quantitative
experiments with an acceptable precision did not succeed, owing to too great dispersion of measured values; much
more professional laboratory equipment than the one we could acquire, as well as more strict measurement conditions
than those we could guarantee, would be necessary.)

In accordance with the generally accepted terminology, the above described complex of phenomena could appropriately be
called the \emph{autothixotropy} of water.

\section{Proposed Explanation of the Phenomena}

Looking for a preliminary interpretation of our observations, a hypothesis based on "ephemeric polymerisation" of the
water seems to be plausible.  The existence of such a weak polymerisation was suspected decades ago, both
defended and denied by experts. If the ephemeric polymerisation of water is the cause of the observed phenomena,
it indicates that dipole water molecules are establishing chains or a network; first as minute complexes, thereafter combining
successively with one another. The structure then becomes more and more dense while oscillating at a certain amplitude on
a scale of molecules.  Such a structure will be relatively fragile, susceptible perhaps even to differences in the
concentration of materials dissolvable in water, at different points inside the vessel. The Brownian motion can be
observed in the case of a conglomerate of molecules having a non-polar character, owing to collisions with molecules of
water oscillating in the established network. A stirring - not too thorough - of the "old"  water seems to preserve
parts of the network, making the subsequent "dipole polymerisation" quicker than it would be in the "fresh" (i.e. well
stirred) water.  Further, the established structure has a certain degree of elasticity. If the water is deaerated by
boiling, no dissolved air (gases) has a disturbing influence either on the developing process or integrity of the
structures; consequently, the phenomenon appears earlier and is more pronounced.

\end{document}